# DEVELOPING SYNTHETIC INDIVIDUAL-LEVEL POPULATION DATASETS: THE CASE OF CONTEXTUALIZING MAPS OF PRIVACY-PRESERVING CENSUS DATA

**Yue Lin[a]\* and Ningchuan Xiao[a]**

[a] Department of Geography, The Ohio State University, Columbus, USA
\* lin.3326@osu.edu

## Introduction

Maps are artifacts that must be interpreted in their social, cultural, and political contexts rather than simply as representations of truth or reality (Harley, 1990, 1991). In demographic mapping, for example, raw data collected from individual members of a population are often modified before further aggregation and visualization due to widespread public concerns about privacy, resulting in maps conveying information that cannot be perceived as geographical facts (Zayatz, 2007; Hawes, 2020). To help map readers better contextualize maps, cartographers are becoming more transparent about details in the cartographic procedure, such as the algorithms involved and certain accuracy metrics about the mapped data, so that the mapmaking process can be reproduced, and readers can explore impacts of the deviation between what is mapped and what is true that arises from this process (Xiao & Armstrong, 2005; Xiao et al., 2007; Kedron et al., 2021). The availability of raw data is essential to ensure full reproducibility for the contextualization of maps; however, many raw data are sensitive and are rarely made public, especially those that contain private information such as the individual-level population data.

The purpose of this paper is to describe the development of a synthetic population dataset that is open and realistic and can be used to facilitate understanding the cartographic process and contextualizing the cartographic artifacts. We first discuss an optimization model that is designed to construct the synthetic population by minimizing the difference between the summarized information of the synthetic populations and the statistics published in census data tables. We then illustrate how the synthetic population dataset can be used to contextualize maps made using privacy-preserving census data. Two counties in Ohio are used as case studies.

## Methods

The synthetic population in an area can be constructed based on the United States Census Summary File 1 (SF1), where each table provides population counts by ethnicity, race, age, housing type, or combinations of these attributes at the census block level (United States Census Bureau, 2011). We begin with a matrix representation of the individual-level population data that needs to be synthesized. Assume that each individual in the dataset has $d$ attributes. A predicate is a tuple consisting of $d' \leq d$ attribute values, and the set of all possible predicates formed by the $d$ attributes is denoted as $S$. The individual-level data is then represented as $X = \{x_{kj}\}$, where $x_{kj}$ is the number of individuals with the $k$-th predicate of $S$ in block $j$ (Figure 1a). The SF1 tables selected are represented in the matrix form of $Y^{(1)}, Y^{(2)}, \ldots, Y^{(n)}$, where $n$ is the number of tables containing population counts by one or more attributes, and each table $Y^{(p)} = \{y_{ij}^{(p)}\}$ has element $y_{ij}^{(p)}$ representing the cell value of row (block) $j$ and column (predicate) $i$ in table $p$ (Figure 1b). Let $W^{(1)}, W^{(2)}, \ldots, W^{(n)}$ be matrices with $W^{(p)} = \{w_{ik}^{(p)}\}$, where element $w_{ik}^{(p)}$ equals one if predicate $i$ for $Y^{(p)}$ is a subset of predicate $k$ for $X$, and zero otherwise. These matrices are referred to as query matrices (Figure 1c).

An optimization problem can be formulated to minimize the sum of squared differences between census data ($Y^{(p)}$) and summary statistics of the synthetic population ($W^{(p)}X$), where each element in $X$ is a decision variable that needs to be determined:

$$min \quad \sum_{p=1}^{n} \left\| W^{(p)}X - Y^{(p)} \right\|^2 \quad (1)$$

$$subject\ to \quad x_{kj} \in Z^*, \forall k, j \quad (2)$$

where constraints (2) ensure integer decision variables.

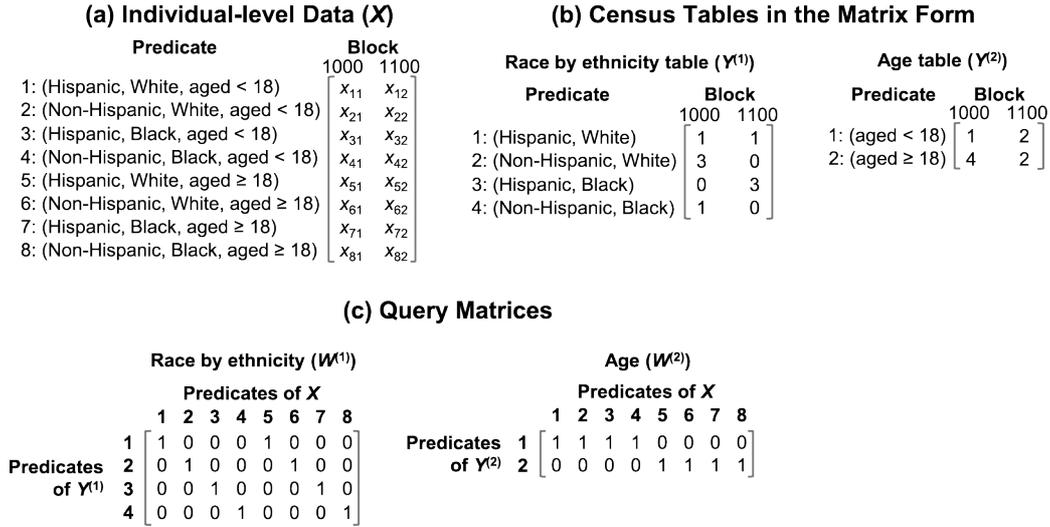

Figure 1: Terminology used in the generation of synthetic populations. Assume a synthetic population dataset for blocks 1000 and 1100 is to be determined, with each individual having three attributes of ethnicity, race, and age. The matrix $X$ in (a) represents such a dataset. Two census tables are selected and are converted to the matrix form, as illustrated in (b). A set of query matrices is defined in (c), and premultiplying a query matrix by $X$ can aggregate the individual-level data to the form of a census data matrix in (b) that enables direct comparison. For example, $W_1X$ returns a race by ethnicity data matrix through aggregating $X$, which can be compared to the census data matrix $Y_1$ to indicate if the summarized information of $X$ is consistent with the census data.

# Results

Franklin and Guernsey counties in Ohio are selected as our study areas because they reflect the diversity of demographic composition and population size in urban and rural areas across the United States. The synthetic population data for both counties have five attributes of housing type, voting age, ethnicity, race, and sex as well as geographic identifiers at the county, census tract, block group, and block levels. More specifically, the data contains 1,163,415 individuals from 22,826 census blocks in Franklin County and 40,088 individuals from 3,768 census blocks in Guernsey County. Histograms illustrating the distributions of block-level population totals in both counties are shown in Figure 2. This dataset, as well as all the code used to generate it, will be made available in a public GitHub repository.

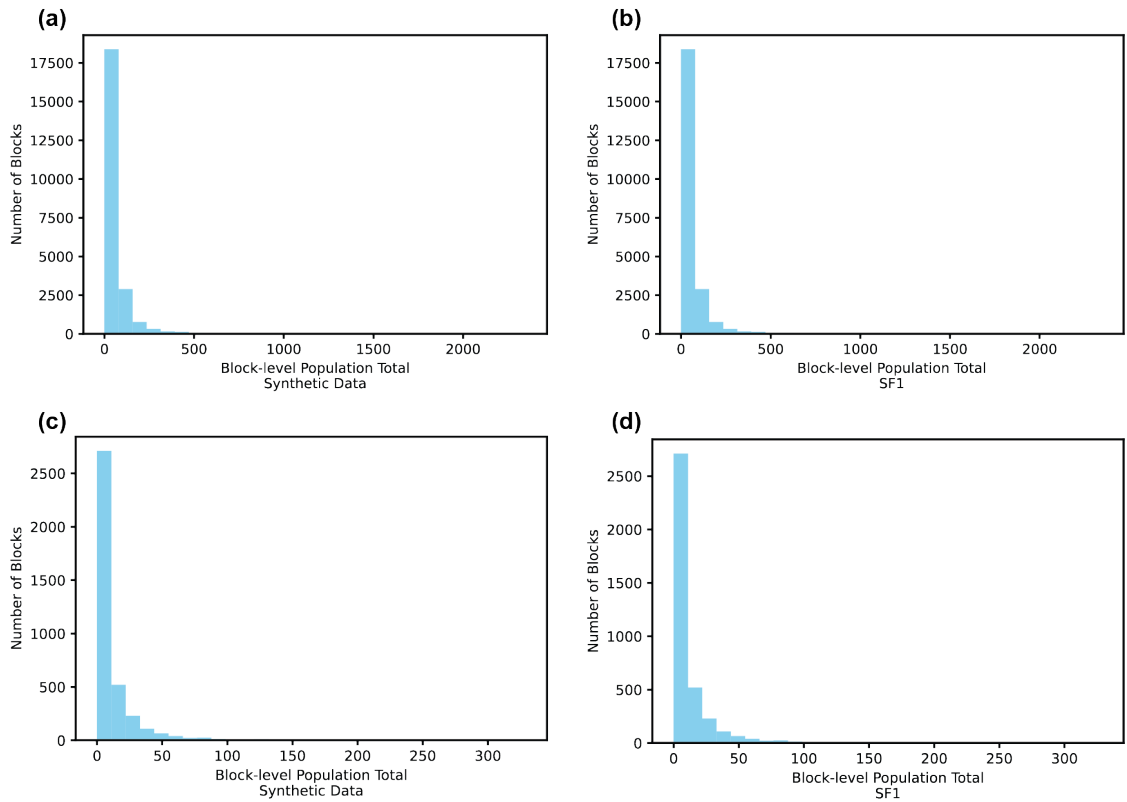

Figure 2: Comparison of block-level population totals in the synthetic data and the SF1 data for Franklin (a, b) and Guernsey (c, d) counties.

## *Technical validation*

Internal (in-sample, or endogenous) and external (out-of-sample, or exogenous) validations are performed to assess the reliability of the synthetic dataset. **For internal validation**, four groups of block-level summary statistics from the synthetic populations are compared to the census SF1 data used in the optimization model. The correlation coefficients (*r*) among the four groups of statistics are all one (Figure 3a, b), indicating that the synthetic populations are well-fitting to the census tabulations. **For external validation**, an external data source known as the American Community Survey Public Use Microdata Sample (ACS PUMS) (United States Census Bureau, 2010) is retrieved for comparison, which is an individual-level dataset that covers a small, representative sample of populations in each county. Specifically, we convert both datasets into their corresponding matrix representations at the county level and calculate the correlation coefficient, which is close to one for both Franklin and Guernsey (Figure 3c, d). This suggests that the synthetic dataset can well represent the real-world populations.

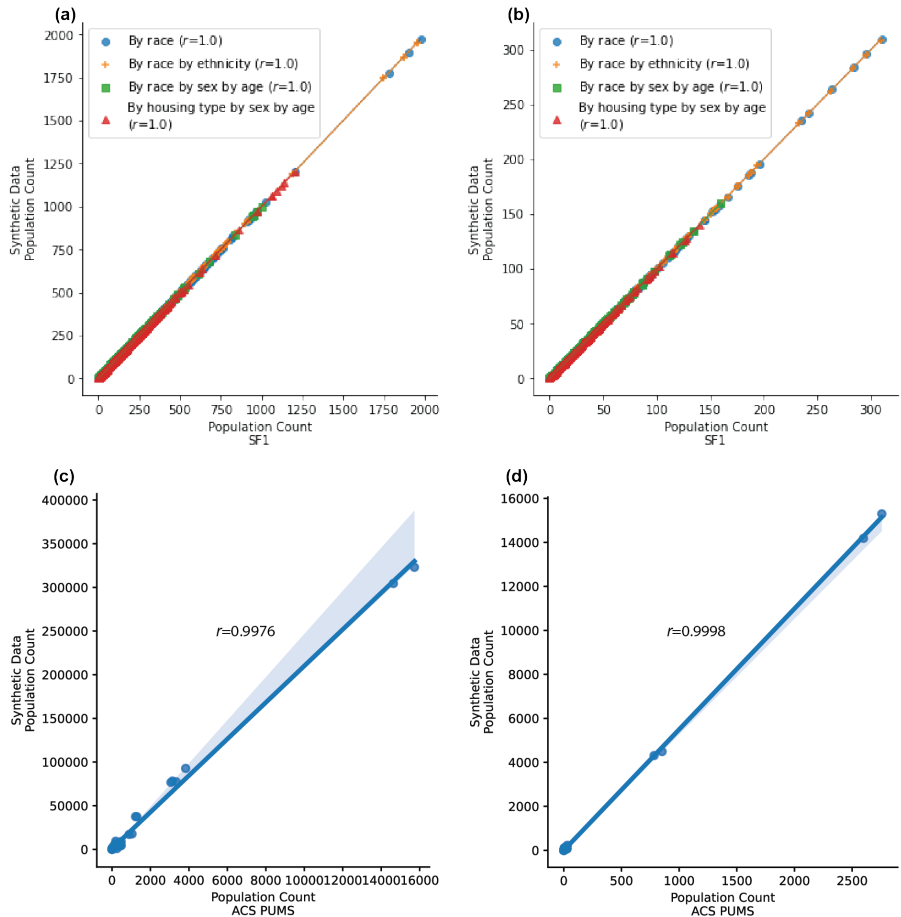

Figure 3: Internal (a, b) and external (c, d) validations of the synthetic dataset, where the correlation coefficients ($r$) are calculated. Each dot in (a) and (b) represents the population counts for a predicate formed by given attribute(s) (colored) within a census block. Each dot in (c) and (d) represents the population counts in the individual-level data matrices at the county level.

*Case study: Contextualizing maps of the 2020 United States Census data*

We provide an example of leveraging the synthetic dataset to contextualize two tract-level racial maps for the percentage of Black or African American created with the 2020 United Census data (Figure 4a, c). In the 2020 United States Census, a privacy protection mechanism known as differential privacy is applied to add statistical noise to the data for protecting individual privacy. To understand how such a mechanism affects what we map using the data, full details of the differential privacy algorithm as well as the original individual-level data should be provided for maps readers to investigate. However, the original data are often not publicly available due to confidentiality constraints, and instead the synthetic population dataset is applied to aid in the contextualization of maps with census data. Specifically, the differential privacy algorithm is implemented on the synthetic data to generate a set of privacy-preserving census data, and the percentage of Black or African American at the tract level is calculated for both the original and the privacy-preserving synthetic data. We compute the symmetric mean absolute percentage error (SMAPE) between these two sets of percentages to indicate the overall error introduced by differential privacy to mapping. The results show that areas with a lower percentage of Black or African American tend to have higher levels of errors introduced to the examined racial maps (Figure 4b, d), which should be taken into consideration in the subsequent use of these maps.

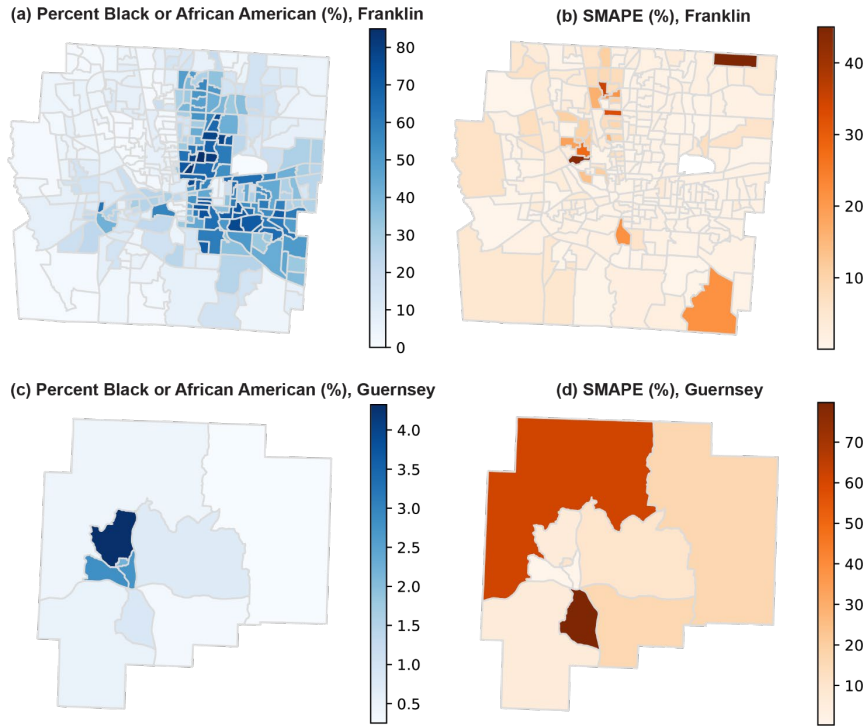

Figure 4: Racial maps (a, c) created with the 2020 Census data and the overall error (b, d).

**Discussion and Conclusion**

We demonstrate in this paper how to generate and use an open and realistic synthetic population dataset to assist in the contextualization of maps, which is especially useful when true data are sensitive and not publicly available. One of the future directions is to expand the scope of this dataset beyond the United States to other regions.